\documentstyle[psfig]{elsart}
\begin{document}
\begin{frontmatter}
\title{Optical Bell Measurement by Fock Filtering}
\author[Pavia,London]{M.G.A. Paris\thanksref{CA}}
\author[London]{M. B. Plenio} \author[London,Oxford]{S. Bose}
\author[London]{D. Jonathan} \author[Pavia]{G. M. D'Ariano}
\address[Pavia]{Theoretical Quantum Optics Group, INFM and Dipartimento di 
Fisica \\ Universit\`a  di Pavia, via  Bassi 6, I-27100 Pavia, ITALY}
\address[London]{Optics Section, Blackett Laboratory, Imperial College, 
London SW7 2BZ, UK} \address[Oxford]{Centre for Quantum Computation, 
Clarendon Laboratory, University of Oxford, Parks Road, Oxford OX1 3PU, UK}
\begin{abstract}
We describe a nonlinear interferometric setup to perform a complete 
optical Bell measurement, i.e. to unambiguously discriminate the four 
polarization-entangled EPR-Bell photon pairs. The scheme is robust 
against detector inefficiency. 
\end{abstract}
\thanks[CA]{Corresponding author. E-mail address: 
{\tt Matteo.Paris@pv.infn.it}} 
\end{frontmatter}
\section{Introduction}
Entanglement and entangled states are fundamental concepts of the new
field of quantum information \cite{review}. They can be exploited
for example in super-dense coding \cite{qdc} or teleportation \cite{tel}
both of which have been demonstrated experimentally. 
The most important example of entangled states is perhaps given by the 
four polarization-entangled Bell states involving two photons
\begin{eqnarray}
|\Psi_\pm\rangle &=& \frac1{\sqrt{2}}\Big[|1001\rangle\pm |0110\rangle\Big]
=\frac1{\sqrt{2}}\Big[a_{\parallel}b_{\perp}\pm a_{\perp}b_{\parallel} 
\Big]^\dag | {\bf 0}\rangle \\
|\Phi_\pm\rangle &=&\frac1{\sqrt{2}}\Big[|1010\rangle\pm |0101\rangle\Big]
=\frac1{\sqrt{2}}\Big[a_{\parallel}b_{\parallel}\pm a_{\perp}b_{\perp} 
\Big]^\dag | {\bf 0}\rangle\label{bellst}\;,
\end{eqnarray}
where $a_{\parallel} (b_{\parallel})$ denotes horizontal polarized and
$a_{\perp} (b_{\perp})$ denotes vertical polarized photons in the two 
possible directions of propagation denoted by $a$ and $b$ (see also 
Fig. \ref{f:setup}). Indeed, these states have
been produced experimentally via the nonlinear process of spontaneous
down-conversion \cite{exp}. In order to actually exploit entanglement,
in the manipulation of quantum information one has to be able to 
distinguish the different maximally entangled states given above. This 
ability is crucial for the unambiguous experimental realization for
example of quantum teleportation \cite{tel}, where an unknown quantum state can 
be exchanged between two parties as long as they share an entangled state, 
and perform the appropriate measurement to discriminate it among the others. 
\par
A simple interferometric setup cannot be of help for the discrimination 
among the four EPR states, as it has been shown recently \cite{nogo,vai} 
that a complete Bell measurement cannot be performed using only 
linear elements. A method to overcome this conclusion has been suggested in
Ref. \cite{emb}, which, however, requires to embed the state of interest 
in a larger Hilbert space, and therefore can be applied only in presence 
of multiple entanglement (entanglement in more than two degrees of freedom).
\section{Bell discrimination}
\begin{figure}\begin{minipage}{11cm}
\psfig{file=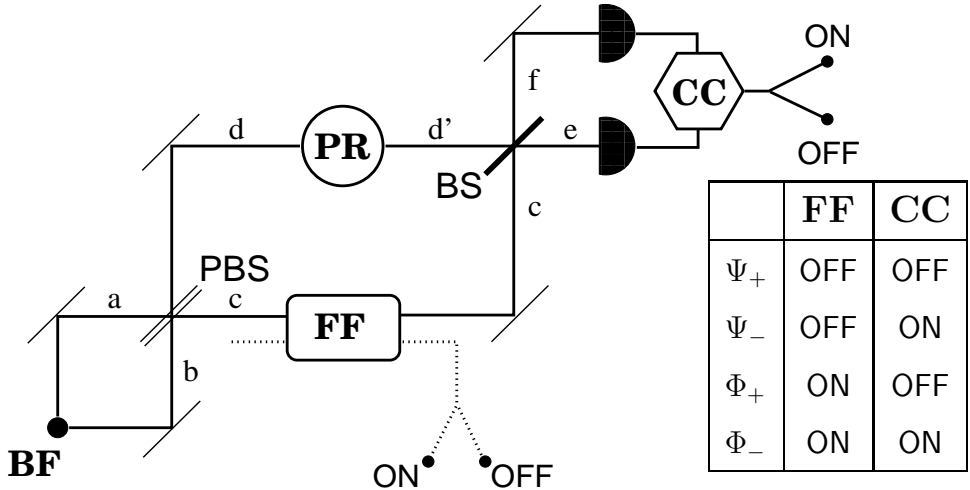,width=12cm}\end{minipage}\hspace{-70pt}
\begin{minipage}{3cm}\vspace{60pt}
\begin{tabular}{|l|c|c|} \hline
& {\large\bf FF} & {\large\bf CC} \\ \hline 
$\Psi_+$ &{\sf OFF} &{\sf OFF}\\ 
$\Psi_-$ &{\sf OFF} &{\sf ON} \\ 
$\Phi_+$ &{\sf ON}&{\sf OFF}\\ 
$\Phi_-$ &{\sf ON}&{\sf ON} \\ \hline  
\end{tabular}\end{minipage}
\caption{Schematic diagram of the nonlinear interferometric setup for the
discrimination of the four Bell states.  One of the four Bell states is
produced at the "Bell Factory" (BF) and then impinges onto a polarizing beam
splitter, whose action is to transmit photons with a fixed polarization (say
vertical) and to reflect photons with the orthogonal one (say horizontal).
Inside the interferometer the polarization of the photons in one arm is
rotated (PR) by a half wave plate , whereas in the other arm 
a non-demolition measurement of the photon number is performed by means 
of a "Fock Filter" (FF) based on Kerr nonlinear interaction. Finally, the 
photons are recombined by an usual, not polarizing, balanced beam splitter 
(BS) and then revealed by a couple of avalanche single-photon 
photo-detectors. A coincidence circuit (CC) tells us whether the photons 
arrived one for each path or both packed in the same one. The field modes 
$c$, $d$, $e$ and $f$ are the Heisenberg evolute of the input
field modes $a$ and $b$. Their explicit expressions are given in the text. 
The inset table describes the reaction of the two measurement stages 
(the Fock Filter FF and the coincidence circuit CC) to the presence of 
the four Bell states respectively.}\label{f:setup}
\end{figure}

In this paper we consider the nonlinear interferometric setup depicted in 
Fig. \ref{f:setup}. Our starting point is a reliable source of optical 
EPR-Bell states, i. e., polarization entangled photon pairs. This is 
usually a birefringent crystal where type-II parametric down-conversion 
transforms an incoming pump photon
into a pair of correlated ordinary and extraordinary photons.
We assume that each pulse (the signal) is prepared in one of the four 
EPR-Bell states, and we want to unambiguously infer from measurements which 
one was actually impinging onto the apparatus. 
\par The signal first enters a polarizing beam splitter, which transmits 
photons with a given polarization (say vertical) and reflects photons
with the orthogonal one (say horizontal). The mode transformations of 
this element is given by (the notation for the field modes refers to Fig.
\ref{f:setup} hereafter) 
\begin{eqnarray}
\left(c_{\parallel}, c_{\perp}, d_{\parallel},d_{\perp} \right) = 
\hat U_{\sc PBS}\left( a_{\parallel}, a_{\perp}, b_{\parallel}, b_{\perp}
\right)\hat U_{\sc PBS}^\dag =
\left(b_{\parallel}, a_{\perp}, a_{\parallel},b_{\perp} \right)
\label{PBSm}\;,
\end{eqnarray}
and the corresponding Schr\"{o}dinger evolution of the 
Bell states is   
\begin{eqnarray}
\hat U_{\sc PBS} \left( \Phi_+ , \Phi_- , \Psi_+ ,  
\Psi_-\right)  = \left(\Phi_+, \Phi_- , \chi_+ ,\chi_- \right) 
\label{PBSs}\;.
\end{eqnarray}
In Eq. (\ref{PBSs}) $\chi_\pm$ are superpositions of states 
with both photons in the same path (arm)
\begin{eqnarray}
|\chi_\pm\rangle &=& \frac1{\sqrt{2}}\Big[|1100\rangle\pm |0011\rangle\Big]
=\frac1{\sqrt{2}}\Big[b_{\parallel}b_{\perp}\pm a_{\parallel} a_{\perp}
\Big]^\dag | {\bf 0}\rangle \label{chis}\;.
\end{eqnarray}
The two sets of states, $\chi_\pm$ and $\Phi_\pm$, can now be 
discriminated by the number of photons travelling in one arm of the
interferometer, which is either zero or two for $\chi_\pm$ and (certainly
only) one for $\Phi_\pm$. Such a discrimination can be performed by means
of a Fock Filter (FF), which is a novel kind of all-optical nonlinear
switch \cite{ff}. Let us postpone the detailed description of the FF to
section 3.
For the moment we assume that it switches on when a single photon
(of any polarization) is present, and does not switch for zero or more than
one photons. As we will see, the FF performs a kind of non-demolition
measurement of the photon number, such that coherence is preserved and
the state after the measurement is still available for further 
manipulations. Indeed, the remaining part of the device should be able to
distinguish phases, namely to discriminate between $\chi_+$ and $\chi_-$, 
or between $\Phi_+$ and $\Phi_-$. For this purpose, first the polarization 
of photons in the second arm is rotated by $\pi/2$ using a half-wave plate, 
thus turning $\Phi_\pm$ into  $\Psi_\pm$ respectively, while leaving 
$\chi_\pm$ untouched. In fact, the transformation induced by the 
polarization rotator reads $\hat U_{PR} = \hat {\sc I} 
\otimes \hat V_{PR}$, where $\hat V_{PR}$ acts only on two modes 
\begin{eqnarray}
\left( d'_{\parallel}, d'_{\perp}\right)=\hat V_{\sc PR} 
\left(d_{\parallel}, d_{\perp} \right)\hat V_{\sc PR}^\dag  
= \left(d_{\perp}, - d_{\parallel}\right) \label{PRm}\;,
\end{eqnarray}
and thus
\begin{eqnarray}
\hat U_{\sc PR} \left(\chi_+,\chi_-, \Phi_+ , \Phi_- \right) = 
\left(\chi_+,\chi_-, \Psi_+ , - \Psi_-\right) \label{PR}\;.
\end{eqnarray}
The two paths are then recombined into a balanced (not polarizing) beam
splitter, whose action on generic field modes $x$ and $y$ is 
described by
\begin{eqnarray}
\hat U_{\sc BS} 
\left(x_{\parallel} , x_{\perp} , y_{\parallel} , y_{\perp}\right)
\hat U_{\sc BS}^\dag =\frac1{\sqrt{2}} 
\left( x_{\parallel} + y_{\parallel}  , 
x_{\perp} + y_{\perp} , x_{\parallel} - y_{\parallel}  , 
x_{\perp} - y_{\perp}\right)
\label{BSm}\;.
\end{eqnarray}
If the transformation (\ref{BSm}) is applied to the field modes $c$ 
and $d'$ we have, using Eqs. (\ref{PRm}) and (\ref{PBSm}),  
\begin{eqnarray}
\left(e_{\parallel} , e_{\perp} , f_{\parallel} , f_{\perp}\right)
&=& \hat U_{\sc BS} \left(c_{\parallel} , c_{\perp} , d'_{\parallel} , 
d'_{\perp}\right) \hat U_{\sc BS}^\dag 
\nonumber \\ &=& \frac1{\sqrt{2}} \left( c_\parallel + d'_\parallel, 
c_\perp + d'_\perp, c_\parallel - d'_\parallel,
c_\perp - d'_\perp \right) \nonumber \\ &=& 
\frac1{\sqrt{2}} \left( b_\parallel + b_\perp, 
a_\perp - a_\parallel, b_\parallel - b_\perp,
a_\perp + a_\parallel \right)
\label{BSMef}\;.
\end{eqnarray}
In terms of the state just before the BS this corresponds to the 
following Schr\"{o}dinger evolution
\begin{eqnarray}
\hat U_{\sc BS} \left( \chi_+ , \chi_- , \Psi_+ , 
\Psi_-\right)  = \left( \chi_+ , \Psi_+ , 
\chi_- , - \Psi_-\right) 
\label{BSs}\;.
\end{eqnarray}
Finally, the photons are measured by single-photons avalanche photo-detectors,
where the last stage of the setup is a coincidence circuit (CC). In fact,
$\Psi_\pm$ correspond to the presence of photons in both channel, whereas 
$\chi_\pm$ are superpositions with both photons in the same path. In terms 
of the states {\em before} the BS, this means that superpositions with the
minus sign ($\chi_-$ and $\Psi_-$) lead to coincident clicks at 
photo-detectors (CC {\sf ON}), whereas superpositions with the plus sign
($\chi_+$ and $\Psi_+$) do not switch on the coincidence circuit.
\par
The whole chain of transformations of our setup  can be summarized by the 
following diagram
\begin{center} \begin{tabular}{ccccccccc}
$\Psi_+$ & $\stackrel{\sc PBS}{\longrightarrow}$ & $\chi_+$ & 
[{\rm FF} {\sf OFF}] & $\stackrel{PR}{\longrightarrow}$ & $\chi_+$ & 
$\stackrel{BS}{\longrightarrow}$ & $\chi_+$ & [{\rm CC} {\sf OFF}] \\
$\Psi_-$ & $\stackrel{\sc PBS}{\longrightarrow}$ & $\chi_-$ & 
[{\rm FF} {\sf OFF}] & $\stackrel{PR}{\longrightarrow}$ & $\chi_-$ & 
$\stackrel{BS}{\longrightarrow}$ & $\Psi_+$ & [{\rm CC} {\sf ON}] \\
$\Phi_+$ & $\stackrel{\sc PBS}{\longrightarrow}$ & $\Phi_+$ & 
[{\rm FF} {\sf ON}] & $\stackrel{PR}{\longrightarrow}$ & $\Psi_+$ & 
$\stackrel{BS}{\longrightarrow}$ & $\chi_-$ & [{\rm CC} {\sf OFF}] \\
$\Phi_-$ & $\stackrel{\sc PBS}{\longrightarrow}$ & $\Phi_-$ & 
[{\rm FF} {\sf ON}] & $\stackrel{PR}{\longrightarrow}$ & $-\Psi_-$ & 
$\stackrel{BS}{\longrightarrow}$ & $\Psi_-$ & [{\rm CC} {\sf ON}] 
\label{chain}\end{tabular}\end{center}
which illustrates the unambiguous discrimination of the four Bell states.
\par
Our scheme is minimal, as it involves only two measurements, and thus
only four possible outcomes, which equals the number of states to be
discriminated.
\section{Fock Filtering}
\begin{figure}[b]
\psfig{file=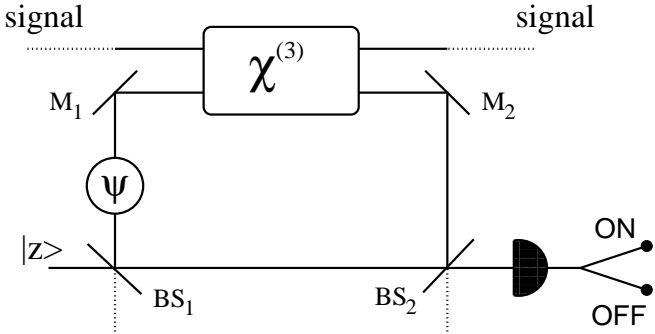,width=9cm}
\caption{Schematic diagram of the Fock Filter. The signal modes are coupled
to the cavity mode by a nonlinear crystal with relevant third-order 
susceptibility $\chi^{(3)}$. The resulting cross-phase modulation imposes
to the cavity mode a phase-shift proportional to the number of photons of the
signal. The ring cavity is built by the mirrors $M_1$ and $M_2$, and by 
the low transmitivity beam splitters $BS_1$ and $BS_2$, whereas $\psi$ 
denotes an externally tunable phase-shift. The cavity is fed by a 
strong coherent state $|z\rangle$, and its output is monitored by an 
avalanche photo-detector.}\label{f:FF}
\end{figure}

The Fock Filter is schematically depicted in Fig. \ref{f:FF}. The signal 
under examination is coupled to a high-Q ring cavity by a nonlinear 
crystal with relevant third-order susceptibility $\chi\equiv\chi^{(3)}$,
which imposes cross-Kerr phase modulation. We assume that the coupling is
independent of polarization, such that the evolution operator of the
Kerr medium is given by 
\begin{eqnarray}
U_K = \exp \left\{- i g (a_{\parallel}^\dag a_{\parallel} + a_{\perp}^\dag 
a_{\perp} )c^\dag c \right\}
\label{uk}\;,
\end{eqnarray}
where $g=\chi t$ is the coupling constant, the $a$'s are the two polarization
modes of the signal, and $c$ describes the cavity mode. Eq. (\ref{uk}) states
that, as a result of the Kerr interaction, the cavity mode is subjected to
a phase-shift proportional to the number of photons passing through the arm
of the interferometer. The FF is complemented by a further, tunable,
phase-shift $\psi$, and operates with the ring cavity fed by a strong coherent
state $|z\rangle$, i.e. a weak laser beam provided by a stable source (the
second port of the cavity is left unexcited).
The input state of the whole device can be written as $|\varphi_{\sc in}
\rangle =  |z\rangle|0\rangle|\nu\rangle$, where 
$|\nu\rangle =\sum_{n_{\perp}n_{\parallel}}
\nu_{n_{\perp}n_{\parallel}}|n_{\perp}\rangle |n_{\parallel}\rangle$ denotes a
generic preparation of the signal mode. The output state is given by
\begin{eqnarray}
|\varphi_{\sc out}\rangle =\sum_{n_{\perp}n_{\parallel}}
\nu_{n_{\perp}n_{\parallel}}
|\sigma_{n_{\perp}+n_{\parallel}}z\rangle
|\kappa_{n_{\perp}+n_{\parallel}}z\rangle
|n_{\perp}\rangle |n_{\parallel}\rangle
\label{ffout}\;,
\end{eqnarray}
where
\begin{eqnarray}
\sigma_n =\frac\tau{1-[1-\tau] e^{i\phi_n}} \qquad
\kappa_n =\frac{\root\of{1-\tau}(e^{i\phi_n}-1)}{1-[1-\tau]e^{i\phi_n}} 
\label{sk}\;,
\end{eqnarray}
are the overall photon-number-dependent transmitivity and reflectivity
of the cavity. In Eq. (\ref{sk}) $\phi_n=\psi-gn$, whereas $\tau$ denotes the
transmitivity of the cavity beam splitters $BS_1$ and $BS_2$. For a good
cavity, (i.e. a cavity with large quality factor) $\tau$ should be quite
small, usual values achievable in quantum optical labs are about $\tau \simeq 
10^{-4} - 10^{-6}$ (losses, due to absorption processes, are about
$10^{-7}$). 
At the cavity output one mode is ignored, whereas the other one is monitored
by an avalanche single-photon photo-detector, which checks whether or not 
photons are present. This kind of {\sf ON/OFF} measurement is described by a
two valued POM 
\begin{eqnarray}
\hat \Pi_{\sf OFF} = \sum_k (1-\eta)^k |k\rangle\langle k| \qquad
\hat \Pi_{\sf ON} = \hat 1 - \hat \Pi_{\sf OFF} \label{onoff}\;,
\end{eqnarray}
$\eta$ being the quantum efficiency of the photo-detector. 
If the state travelling through the interferometer is either $\Phi_+$ or
$\Phi_-$ we have $|\varphi_{\sc in}\rangle = 
|z\rangle|0\rangle|\Phi_\pm \rangle$ and thus
\begin{eqnarray}
|\varphi_{\sc out}\rangle = \frac1{\sqrt{2}}\left[
|\sigma_1 z\rangle|\kappa_1 z\rangle|1010\rangle
\pm |\sigma_1 z\rangle|\kappa_1 z\rangle|0101\rangle\right]=
|\sigma_1 z\rangle|\kappa_1 z\rangle|\Phi_\pm\rangle
\label{phiout}\;.
\end{eqnarray}
The probability of having a click is given by 
\begin{eqnarray}
P({\sf ON}| \Phi_\pm) = \hbox{Tr} \left\{|\varphi_{\sc out}\rangle\langle 
\varphi_{\sc out}| \: \hat\Pi_{\sf ON}\right\}= 1-\exp \left( - \eta
|\sigma_1|^2 |z|^2\right)
\label{prob1}\;,
\end{eqnarray}
whereas the conditional output signal state, after a click has been actually
registered, turns out to be
\begin{eqnarray}
\hat \nu _{\sc out} ({\sf ON}| \Phi_\pm) = \frac1{P({\sf ON}| \Phi_\pm)} 
\: \hbox{Tr}_{cavity} \left\{|\varphi_{\sc out}\rangle\langle 
\varphi_{\sc out}| \: \hat\Pi_{\sf ON}\right\} = |\Phi_\pm\rangle\langle\Phi_\pm|
\label{out1}\;.
\end{eqnarray}
By setting $\psi=g$ and for $\eta |z|^2 \gg 1$ we have $P({\sf ON}|
\Phi_\pm)\simeq 1$.
On the other hand, if the signal is either $\chi_+$ or $\chi_-$ we have
$|\varphi_{\sc in}\rangle =  |z\rangle|0\rangle|\chi_\pm \rangle$ and 
\begin{eqnarray}
|\varphi_{\sc out}\rangle &=& \frac1{\sqrt{2}}\left[
|\sigma_2 z\rangle|\kappa_2 z\rangle|1100\rangle
\pm |\sigma_0 z\rangle|\kappa_0 z\rangle|0011\rangle\right] \nonumber \\
&=& \frac1{\sqrt{2}}\left[
|\sigma z\rangle|\kappa z\rangle|1100\rangle
\pm |\sigma^* z\rangle|\kappa^* z\rangle|0011\rangle\right] 
\label{chiout}\;,
\end{eqnarray}
where the second equality comes from the fact that by setting $\psi=g$ we have
$\sigma_0^*=\sigma_2\equiv \sigma$. Finally, from Eqs. 
(\ref{onoff}) and (\ref{chiout}) we have
\begin{eqnarray}
P({\sf OFF}| \chi_\pm) &=& \exp \left( - \eta
|\sigma|^2 |z|^2\right) \\ 
\hat \nu _{\sc out} ({\sf OFF}| \chi_\pm) &=& |\chi_\pm\rangle\langle\chi_\pm|
\label{outprob2}\;.
\end{eqnarray}
For small $g$ and $\tau$ we have $|\sigma|^2\simeq (\tau/g)^2$. Therefore,
for $\tau \ll g$ we have  $P({\sf OFF}| \chi_\pm)\simeq 1$. 
This means that a click at the Fock Filter unambiguously implies that either
$\Phi_+$ or $\Phi_-$ was travelling through the interferometer, where having
no click indicates the passage of either $\chi_+$ or $\chi_-$. Remarkably, 
the measurement is quite robust against detector inefficiency (as only the
product $\eta |z|^2$ is relevant) and does not destroy coherence. For both the 
possible outcomes (either {\sf ON} or {\sf OFF}) the state after the
measurement remains unaffected, and is still available for further
manipulations. 
\section{Conclusions}
In conclusion, we have described an interferometric setup to perform 
a complete optical Bell measurement. It consists of a Mach-Zehnder
interferometer with the first beam splitter a polarizing one, and 
the second a normal one, and where inside the interferometer a 
non-demolition photon number measurement is performed by the Fock 
filtering technique. The resulting scheme is robust  against detectors 
inefficiency, and provides a reliable method to unambiguously discriminate 
among the four polarization-entangled EPR-Bell photon pairs. 
\ack
We would like to thank D. Bouwmeester for useful comments. 
This work was supported by CNR and NATO through the Advanced Fellowship
program 1998, the EPSRC, The Leverhulme Trust, the INFM, the Brazilian agency 
Conselho Nacional de Desenvolvimento Cientifico e Tecnologico (CNPQ), 
the ORS Awards Scheme, the Inlaks Foundation and the European Union. 


\begin{thebibliography}{99}
\bibitem{review} M.B. Plenio and V. Vedral, Cont. Physics {\bf 39}, 431 (1998). 
\bibitem{qdc} C. H. Bennett, S. J. Wiesner, Phys. Rev. Lett. {\bf 69}, 2881
(1992); K. Mattle, H. Weinfurter, P. G. Kwiat, A. Zeilinger, Phys. Rev. Lett.
{\bf 76}, 1895 (1996).
\bibitem{tel} C.H. Bennett, G. Brassard, C. Crepeau, R. Jozsa, A. Peres, and 
W.K. Wootters, Phys. Rev. Lett. {\bf 70}, 1895 (1993); D. Boschi, S. Branca, 
F. De Martini, L. Hardy, S. Popescu, Phys. Rev. Lett. {\bf 80}, 1121 (1998); 
S. L. Braunstein, H. J. Kimble, Phys. Rev. Lett. {\bf 80}, 869 (1998); 
D. Bouwmeester, J.W. Pan, K. Mattle, M. Eibl, H. Weinfurter, A. Zeilinger, 
Nature {\bf 390}, 575 (1997).
\bibitem{exp} P. G. Kwiat, K. Mattle, H. Weinfurter, A. Zeilinger, A. V.
Sergienko, Y. Shih, Phys. Rev. Lett, {\bf 75}, 4337 (1995).
\bibitem{nogo} N. L\"{u}tkenhaus, J. Calsamiglia, K.-A. Suominen, Phys. Rev.
A{\bf 59}, 3295 (1999).
\bibitem{vai} L. Vaidman, N. Yoran, Phys. Rev. A{\bf 59}, 116 (1999).
\bibitem{emb} P. G. Kwiat, H. Weinfurter, Phys. Rev. A{\bf 58}, R2623 (1998).
\bibitem{ff} G. M. D'Ariano, L. Maccone, M. G. A. Paris, M. F. Sacchi, 
Acta Phys. Slov. {\bf 49}, 659 (1999); see also quant-ph 9906077.
\end{thebibliography}
\end{document}